\documentclass[useAMS,usenatbib]{mn2e}
\usepackage{epsfig}
\usepackage{amsmath}

\newcommand{\be}{\begin{equation}}
\newcommand{\beq}{\begin{equation}}
\newcommand{\ba}{\begin{eqnarray}}
\newcommand{\ee}{\end{equation}}
\newcommand{\eeq}{\end{equation}}
\newcommand{\ea}{\end{eqnarray}}

\newcommand{\apj}{ApJ}
\newcommand{\apjl}{ApJL}
\newcommand{\mnras}{MNRAS}
\newcommand{\aj}{AJ}
\newcommand{\apjs}{ApJS}
\newcommand{\nat}{{\it Nature}}

\def\lsim{~\rlap{$<$}{\lower 1.0ex\hbox{$\sim$}}}

\def\gsim{~\rlap{$>$}{\lower 1.0ex\hbox{$\sim$}}}

\title[Clustering of High Redshift Quasars]{Evidence for Merger-Driven
Activity in the Clustering of High Redshift Quasars}

\author[Wyithe \& Loeb]{J. Stuart B. Wyithe$^1$ and Abraham Loeb$^2$\\$^1$
School of Physics, University of Melbourne, Parkville, Victoria,
Australia\\$^2$ Harvard-Smithsonian Center for Astrophysics, 60 Garden St.,
Cambridge, MA 02138\\Email: swyithe@unimelb.edu.au,
loeb@cfa.harvard.edu}

\begin{document}


\maketitle

\label{firstpage}
\begin{abstract}

\noindent 
Recently, a very large clustering length has been measured for quasars
at a redshift of $z\sim4$. In combination with the observed quasar
luminosity function we assess the implications of this clustering for
the relationship between quasar luminosity and dark matter halo
mass. Our analysis allows for non-linearity and finite scatter in the
relation between quasar luminosity and halo mass, as well as a
luminosity dependent quasar lifetime. The additional novel ingredient
in our modelling is the allowance for an excess in the observed bias
over the underlying halo bias owing to the merger driven nature of
quasar activity. We find that the observations of clustering and
luminosity function can be explained only if both of the following
conditions hold: {\it (i)} The luminosity to halo mass ratio increases
with halo mass; {\it (ii)} The observed clustering amplitude is in
excess of that expected solely from halo bias. The latter result is
statistically significant at the 99\% level. Taken together, the
observations provide compelling evidence for merger driven quasar
activity, with a black-hole growth that is limited by feedback. In
difference from previous analyses, we show that there could be scatter
in the luminosity halo mass relation of up to 1 dex, and that quasar
clustering can not be used to estimate the quasar lifetime.

\end{abstract}


\section{Introduction}

The Sloan Digital Sky Survey (SDSS; York et al.~2000) and the 2dF
quasar redshift survey (Croom et al.~2001a) have measured redshifts
for large samples of quasars, and determined their luminosity function
over a broad section of cosmic history (Boyle et al.~2000; Richards et
al.~2006). These surveys have also been used to constrain the
clustering properties of quasars (e.g. Croom et al.~2001b; Croom et
al.~2002; Croom et al. 2005; Porciani \& Norberg~2006; da Angela et
al.~2008; Padmanabhan et al.~2008), including the variation of
clustering length with redshift and luminosity.  In the local Universe
quasars have clustering statistics similar to optically selected
galaxies, with a clustering length $R_0\approx 8$Mpc. The clustering
length increases towards high redshift, but is only weakly dependent
on luminosity. This has been interpreted as evidence for a model in
which the quasar luminosity takes on a broad range of values at
different stages of its evolution (Lidz et al.~2006).

In addition to the large quasar samples below $z\sim3$, the Sloan
Digital Sky Survey (SDSS) has discovered luminous quasars at redshifts
as high as $z\sim6.4$, i.e., when the universe was less than a billion
years old (Fan et al.~2001a,b; Fan et al.~2003; Fan et al.~2006). The
supermassive black-holes (SMBH) powering these quasars have masses of
$\ga 10^9M_\odot$. Recently, clustering measurements of quasars have
been extended out to redshifts beyond $z\sim4$ using the Fifth Data
release of the SDSS (Shen et al.~2007). Quasars are shown to be
significantly more clustered at high redshift relative to the more
local samples, with values of $R_0\approx 12$Mpc and $R_0\approx
17$Mpc at $z\sim3$ and $z>3.5$ respectively (Shen et al.~2007). White,
Martini \& Cohn~(2008) have recently shown that this large clustering
length can be associated with the observation of a very large
clustering bias of $b=14.2\pm1.4$ for quasars at $z\sim4$.

It was argued by Martini \& Weinberg~(2001) and Haiman \& Hui~(2001),
that the quasar correlation length can be used to infer the typical mass of
dark matter halos in which quasars reside. One may then derive
the quasar duty-cycle by comparing the number density of quasars with
the density of host dark matter halos. The quasar lifetime follows
from the product of the duty-cycle and the time that the dark-matter
halo spends in between major mergers (although there is a degeneracy
between the lifetime and the quasar occupation fraction or beaming).
Results from low redshift clustering have suggested quasar lifetimes
of $t_{\rm q}\sim10^6$--$10^7$ years, consistent with the values
determined by other methods (see Martini~2003 for a review). More
recently, Shen et al.~(2007) have applied this analysis to their
measurements of quasar clustering at high redshift. They infer
lifetimes of $t_{\rm q}\sim3\times10^7$--$6\times10^8$ years for
quasars at $z>3.5$.

Using techniques similar to those employed at low redshift, this
result has been used by White et al.~(2008) to argue that the
scatter in the relation between quasar luminosity and halo mass must
be smaller than 0.3dex. This small scatter poses a problem for
theories of quasar growth and formation. The tightest local relation
is observed between the black hole mass and bulge velocity
dispersion, also with a scatter of $\sim 0.3$ dex. However, the relation
between quasar luminosity and halo mass must have several additional
sources of scatter, including in the relation between halo mass and
velocity dispersion, and between black hole mass and quasar
luminosity. It is therefore difficult to understand how the luminosity
-- halo mass relation could be as tight as the black-hole mass --
velocity dispersion relation.

The conclusion of White et al.~(2008) arises because an increased
scatter would tend to bring large numbers of low bias halos (which are
more common) into a sample of halos at fixed quasar luminosity. The
required small scatter therefore depends crucially on the assumption
that the value of the observed bias truly reflects the actual bias of
the quasar host galaxies. However additional bias may exist beyond the usual
halo bias for systems which are triggered by mergers, although the
magnitude (and even the sign) of the effect are still being debated
(e.g., Kolatt et al.~1999; Gottlober et al.~2002; Kauffmann \&
Haehnelt~2002; Percival et al.~2003; Scannapieco \& Thacker~2003; Gao,
Springel \& White~2005; Wetzel et al.~2007). 
Interpreting the results of simulations of
the effect of mergers on bias is complicated by the small range of
redshift and mass available in different studies. Furlanetto \&
Kamionkowski~(2006) attempted to describe the clustering bias of
merging systems more generally based on an analytic model. They
concluded that a peak-background split approach within the extended
Press-Schechter formalism cannot be used to calculate the merger bias
because the large scale density field does not enter the calculation
of merger rates using this approach. Instead, Furlanetto \&
Kamionkowski~(2006) calculated the merger bias in models where the
merger rate per halo is assumed to scale with the number densities of
neighboring halos. Since this model does not account for the large
scale density in the merger rate itself (but only in the halo
densities), Furlanetto \& Kamionkowski~(2006) argue that it can only
be considered qualitative.
However, the model suggests that merging halos are significantly more
clustered than isolated halos, by a factor of $\sim1.5$ for massive
halos at $z\sim3$. Furlanetto \& Kamionkowski~(2006) also proposed a
simple model in which close pairs are assumed to be merging systems,
and compute the bias of those based on probabilities of separation
based on clustering statistics in a variety of models, with similar
results.

The possibility that the observed bias exceeds the average halo bias
at a particular halo mass is not restricted to scenarios where
observed objects are merger driven. A separate, but related issue
concerns halo formation bias. Here sub-samples of dark matter halos of
fixed mass and redshift are shown to have clustering statistics that
depend on their formation history (Gao, Springel \& White~2005; Croton,
Gao \& White~2007; Wetzel et al.~2007). For example, if the existence of
luminous quasars required that the host halo be older than a
particular minimum age, then this sub-sample of halos would have a
clustering bias in excess of the halo population as a whole.

Since luminous quasars play an important role in the evolution of
massive galaxies at high redshifts, a variety of models have been
proposed to explain the high redshift luminosity function of quasars
(e.g. Haiman \& Loeb~1998; Haehnelt, Natarajan \& Rees~1998; Kauffmann
\& Haehnelt~2000; Volonteri, Haardt \& Madau~2003; Wyithe \&
Loeb~2003; Di~Matteo, Springel \& Hernquist~2005; Hopkins \&
Hernquist~2008; Hopkins et al.~2008; Li et al.~2007). The majority of
these models assume that major mergers drive the quasar activity,
implying that the observed bias should not simply reflect the bias
based on the host halo mass.

In this paper we investigate the evidence for an additional bias, 
beyond the spatial clustering of dark matter halos, 
in the clustering data of high redshift quasars. Wyithe \&
Padmanabhan~(2006) showed that different models are able to reproduce
the data on the high redshift quasar luminosity function, implying a
degeneracy among their input physical parameters. For this reason, we
will not attempt to model physical processes such as SMBH growth and
feedback, but instead adopt the more general approach of
parameterising relations like the correlation between SMBH mass and
halo mass.  We are interested in the full range of allowed parameters
as well as in which sets of parameters can be excluded, rather than in
a particular set of parameters that is able to describe the data. This
approach allows us to isolate the relationships that are constrained
by the data, and to provide robust observational input for future
theoretical modeling. In particular, the contribution of mergers to
the observed clustering bias could be large, although its value is
theoretically uncertain. We therefore treat this contribution as a
free parameter, and attempt to constrain its value.

In our numerical examples we adopt the standard set of cosmological
parameters ~(Komatsu et al.~2008), with values of $\Omega_{\rm
m}=0.24$, $\Omega_{\rm b}=0.046$ and $\Omega_Q=0.72$ for the matter,
baryon, and dark energy fractional density respectively, and $h=0.70$,
for the dimensionless Hubble constant. We assume $\sigma_{8}= 0.82$
for the {\it rms} amplitude of the density field fluctuations within
spheres of radius $8h^{-1}$Mpc linearly extrapolated to $z=0$, and a
power-law slope for the primordial density power-spectrum of $n_{\rm
s}=0.96$.

\section{Model}

Quasars are known to have a finite distribution of Eddington ratios
(Kollmeier et al.~2006), $\eta\equiv l/L_{\rm Edd}$, where $l$ and
$L_{\rm Edd}$ are the quasar and Eddington luminosities respectively,
and are thought to have a light-curve that varies across a large range
of luminosity during the quasar lifetime (Hopkins \&
Hernquist~2008). We can express this lightcurve as $l(t,M_{\rm 
  bh})\propto \eta(t) M_{\rm bh}$, where $M_{\rm bh}$ is the black
hole mass and $\eta$ is a dimensionless function of time. Consider now
a population of quasars at a fixed instant in time. We expect to
observe a range of quasar luminosity $l$ at fixed host halo mass
$M$ because different quasars will be at a different phase of
their light curve. We first compute the distribution of luminosity $l$
at fixed halo mass and fixed black hole mass
\begin{equation}
\left.\frac{dP}{dl}\right|_{M,M_{\rm bh}}\propto
\frac{dt}{dl}\frac{dP_{\rm prior}}{dt}\propto \left(M_{\rm bh}
\frac{d\eta}{dt}\right)^{-1}{\Theta(t_{\rm lt})\over t_{\rm lt}}.
\end{equation}
Here, $({dP_{\rm prior}}/{dt})=\Theta(t_{\rm lt})/t_{\rm lt}$ (where
$\Theta$ is the Heaviside Step function), is constant during the quasar
lifetime ($0<t<t_{\rm lt}$), and zero at other times. The distribution
of $l$ at fixed halo mass is then
\begin{equation}
\left.\frac{dP}{dl}\right|_{M} \propto \int dM_{\rm bh}\left. \frac{dP}{dM_{\rm bh}}\right|_{M} \left.\frac{dP}{dl}\right|_{M,M_{\rm bh}},
\end{equation}
where $\left.\frac{dP}{dM_{\rm bh}}\right|_{M}$ is the
probability distribution for the black-hole mass which depends on halo
mass. The mean of the logarithm of the luminosity at halo mass
$M$ can now be calculated from
\begin{equation}
(\log{L})(M) = \int d\log{l}\, \log{l} \left.\frac{dP}{d\log{l}}\right|_{M}, 
\end{equation}
while the variance (in dex) can be calculated as
\begin{equation}
\Delta^2(M) = \int d\log{l} (\log{l}-{\log{L}})^2
\left.\frac{dP}{d\log{l}}\right|_{M}.
\end{equation}

In this paper we do not use the above equations to compute the
relation between the quasar luminosity, $L$, and host halo mass,
$M$. Rather, we suggest a parametrised form based on the above
definitions and constrain the parameters of the $L$--$M$ relation
using the available data for high redshift quasars. We begin by
parameterising the relation between the mean quasar luminosity,
$L$, and host halo mass, $M$, as
\begin{equation}
\frac{L}{L_0} = \left(\frac{M}{M_0}\right)^{\gamma},
\end{equation}
where $L_0$ and $M_{0}$ are normalisation constants.  We then suppose
that this relation has an intrinsic scatter in luminosity $l$ at fixed
halo mass $M$ of $\Delta$. For simplicity, we assume that the
distribution $dP/dl|_{M}$ is log-normal and express its
scatter in dex. Re-writing the mean relation as
\begin{equation}
\label{M-L}
M = M_o\left(\frac{L}{L_0}\right)^{\frac{1}{\gamma}},
\end{equation}
we then specify the scatter in the $M$--$L$ relation in terms
of a probability distribution for halo mass $m$ at fixed luminosity $L$ i.e.
\begin{equation}
\left.\frac{dP}{d\log{m}}\right|_{L} = \frac{1}{\sqrt{2\pi}(\Delta/\gamma)}\exp{\left[\frac{(\log{m}-\log{M})^2}{2(\Delta/\gamma)^2}\right]}.
\end{equation} 

We next specify a model for the luminosity function of quasars
(i.e. the number density of quasars per unit luminosity with
luminosity $L$)
\begin{equation}
\frac{dn}{dL} = \int d\log{m} \frac{dn}{dm}
\left.\frac{dm}{dL}\right|_{L,M_0} \left.\frac{dP}{d\log{m}}\right|_{L} f_{\rm
duty}\left(\frac{L}{L_0}\right)^\alpha,
\label{Eq:8}
\end{equation}
where $\frac{dn}{dM}$ is the Sheth-Tormen~(1999) mass-function for dark
matter halos of mass $M$, and the derivative
\begin{equation}
\frac{dm}{dL} = \frac{dM}{dL}\frac{dm}{dM} = \frac{m}{M}\frac{M_0}{L_0}\frac{1}{\gamma}\left(\frac{L}{L_0}\right)^{\frac{1}{\gamma}-1}.
\end{equation}
The duty-cycle $f_{\rm duty}=t_{\rm lt}/t_H$, where $t_H$ is the
Hubble time at redshift $z$, is the fraction of halos that have a
quasar in a luminous phase, and the last factor in equation
(\ref{Eq:8}) accounts for the possibility that the quasar lifetime
$t_{\rm lt}$ is dependent on black hole mass\footnote{In this paper we
only consider the luminosity function and its slope at a single value
of luminosity. While we have quoted a power-law form for the
dependence of duty cycle on luminosity we note that this should be
thought of as a logarithmic slope at luminosity $L$. In general the
value of $\alpha$ must be luminosity dependent or else the duty-cycle
would exceed unity for some values of luminosity. }. The luminosity
function has an associated logarithmic slope
\begin{equation}
\beta(L)\equiv\frac{d}{d\log{L}}\left(\log{\frac{dn}{dL}}\right).
\end{equation}
We also compute the density of quasars above a limiting luminosity
$L$,
\begin{equation}
n(>L) = \int_L^{\infty} dL^\prime \frac{dn}{dL^\prime}.
\end{equation}
The halo bias of quasars with luminosity $L$ is given by,
\begin{equation}
\bar{b}(L) = \frac{\int d\log{m} \, b(m) \frac{dn}{dm} \left.\frac{dP}{d\log{m}}\right|_{L}}{\int d\log{m} \frac{dn}{dm} \left.\frac{dP}{d\log{m}}\right|_{L}},
\end{equation}
where $b(m)$ is the halo bias of a halo with mass $m$, which is computed using the fitting formula of Sheth, Mo \& Tormen~(1999). This bias can
be used to calculate the halo bias of a sample of quasars with
luminosities above a limiting value.
\begin{equation}
\langle b\rangle = \frac{\int_L^{\infty} dL^\prime \, \bar{b}(L^\prime)\,\frac{dn}{dL^\prime}}{\int_L^{\infty} dL^\prime\, \frac{dn}{dL^\prime}.}
\end{equation}
As noted in \S 1, this average halo bias may underestimate
the observed quasar bias if quasars are triggered by galaxy mergers
(Furlanetto \& Kamionkowski 2006). To account for this possibility, we
introduce a free parameter $F$, defined as the ratio of the observed
bias $b_{\rm obs}$ to the halo bias $\langle b\rangle$,
\begin{equation}
F\equiv\frac{b_{\rm obs}}{\langle b\rangle}.
\end{equation}

Note that the value of the black-hole mass does not enter our
calculations. Thus, our conclusions are insensitive to assumptions
regarding the Eddington ratio and the relation between black-hole mass
and halo mass. Conversely, this means that our model is not able to
address issues surrounding these relations. Nevertheless, as we show
below, this simplicity does allow our model to reach some strong
conclusions regarding the relationship between the observed quasar
clustering and the host halo mass.

Before proceeding to analyse the implications of existing
observational data, we note that the detailed results are sensitive to
the form of the halo mass function. Our fiducial calculations use the
Sheth-Tormen form. However in \S~\ref{systematic} we also present
results using the corresponding Press-Schechter formulation to assess
the level of theoretical uncertainty in the results.

\section{Comparison with observations}

The quantities $n(>L)$, $\beta$ and $b_{\rm obs}$ are observed
properties of the quasar population, with values of
$n(>L)\sim0.7\times10^{-7}$Mpc$^{-1}$, $\beta=-2.58\pm0.23$ and
$b_{\rm obs}=14.2\pm1.4$, respectively.  The quasar sample from which
the clustering and corresponding density data were obtained was
described in Shen et al.~(2007). The bias and density for quasars at
$z\sim 4$ were discussed for this sample in White, Martini \&
Cohn~(2008), and we use their calculated bias in this paper. The
luminosity function slope is taken from the quasar luminosity function
analysis of Fan et al.~(2001b), for which the mean quasar redshift and
absolute AB magnitude at $1450$\AA were $z=4.3$ and $M_{1450}\approx -26$,
respectively. These quantities can be compared to the theoretical
expectations described above, and hence used to constrain the free
parameters, $f_{\rm duty}$, $\alpha$, $\Delta$, $\gamma$ and $F$.

\subsection{Results}
\label{Results}

\begin{figure*}
\centerline{\epsfxsize=5.in \epsfbox{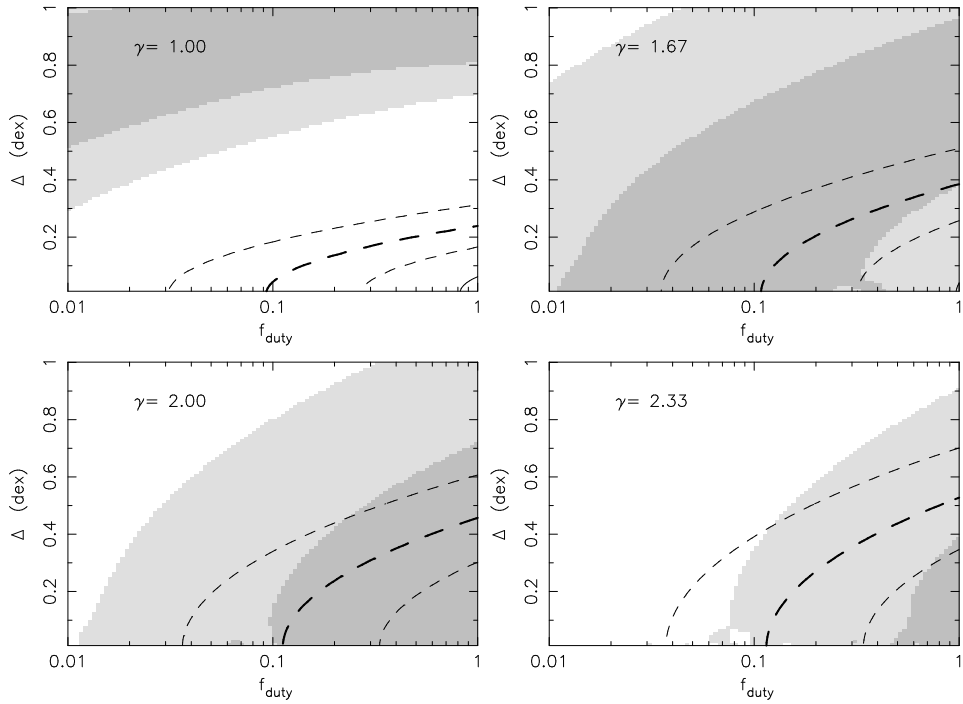}} 
\caption{\textit{Solid and dashed lines:} Contours of joint likelihood
for pairs of $\Delta$ and $f_{\rm duty}$ given the observed bias. The
solid and dashed contours correspond to values of $F=1.0$ and $F=1.5$,
respectively. For $F=1$, the contour is drawn at the 2-$\sigma$ lower
limit of $\langle b\rangle=11.4$.  For $F=1.5$, the contours are drawn
at $\langle b\rangle=(14.2\pm1.4)/F$. In each case the luminosity $L$
corresponds to the observed density of quasars. The grey-scale shows
the corresponding joint likelihood given the constraint of the
observed luminosity function slope $\beta = -2.58\pm0.23$. The dark
and light grey show the regions where the predicted slope is within 1
and 2-sigma of the best fit value. Four examples are shown for
different values of $\gamma$, with $\alpha=0$ in all cases. }
\label{fig1}
\end{figure*}

\begin{figure*}
\centerline{\epsfxsize=5.in \epsfbox{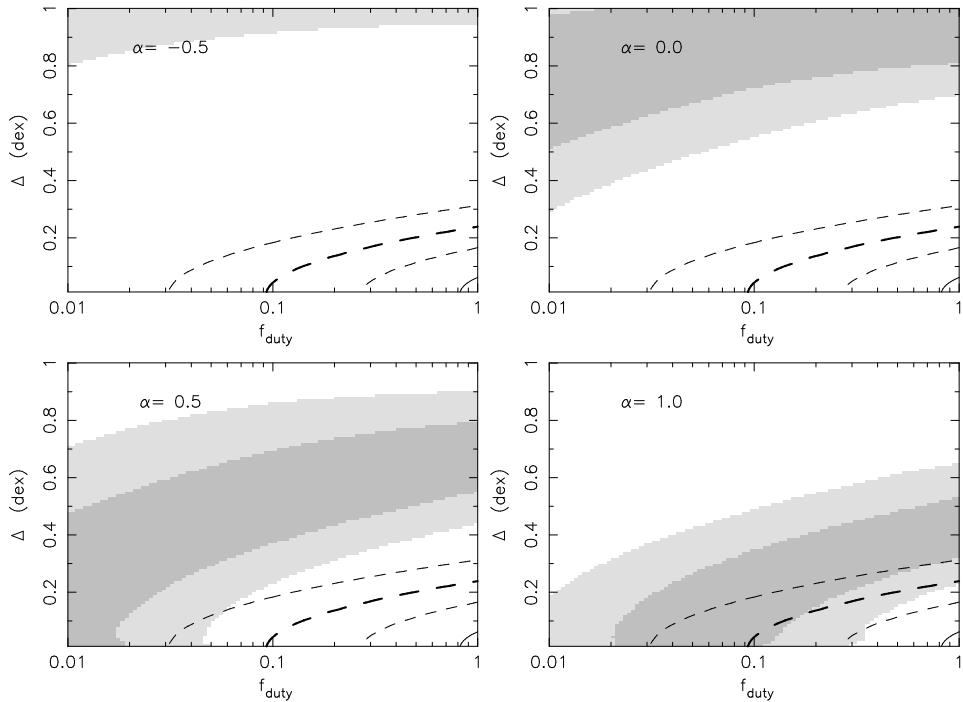}} 
\caption{As per Figure~\ref{fig1}, but for different values of $\alpha$, with $\gamma=1$ in each case.}
\label{fig2}
\end{figure*}

We begin by illustrating the dependencies of the observables on the
available free parameters. In Figures~\ref{fig1} and \ref{fig2} we
show contours of constant $\langle b\rangle/F$ in the plane of $f_{\rm
  duty}$ and $\Delta$. Since the halo bias is related to the observed
bias through an unknown constant $F$, we begin by showing two cases
for illustration. In the first we choose $F=1$, which is the standard
model employed to link clustering statistics to halo mass and quasar
lifetime in previous studies (Martini \& Weinberg~2001; Haiman \&
Hui~2001; White, Martini \& Cohn~2008). In the second, we adopt
$F=1.5$, which is approximately the factor by which the observed bias
exceeds the halo bias in the models of Furlanetto \& Kamionkowski
(2006). In Figures~\ref{fig1} and \ref{fig2} the solid and dashed
contours are plotted for values of $F=1.0$ and $F=1.5$,
respectively. We find that there are no combinations of $f_{\rm duty}$
and $\Delta$ that produce the best fit value of the observed bias for
$F=1$ (see also White, Martini \& Cohn~2008). In this case we draw a contour
(solid line) at the 2-$\sigma$ lower limit of $\langle
b\rangle=(14.2\pm1.4)/F=14.2-2.8=11.4$. In the second case with
$F=1.5$, the contours (dashed lines) are drawn to correspond to the
observed bias $\langle b\rangle=(14.2\pm1.4)/F$.

In constructing these contours, for each combination of $f_{\rm duty}$
and $\Delta$ the luminosity $L$ is chosen so that the luminosity
function reproduces the observed density of quasars. The duty cycle on
the $x$-axis corresponds to quasars with this luminosity $L$.  For
fixed values of $F$, $\gamma$ and $\alpha$, the contours in
Figures~\ref{fig1} and \ref{fig2} enclose the acceptable combinations
of $f_{\rm duty}$ and $\Delta$, given the constraint of the observed
bias. We find that smaller values of duty-cycle permit smaller values
of intrinsic scatter in the $L-M$ relation.  The two figures separate
the effect of the parameters $\alpha$ and $\gamma$ on the derived
constraints. In Figure~\ref{fig1} we set $\alpha=0$ and vary the value
of $\gamma$. In Figure~\ref{fig2}, we fix the value of $\gamma=1$ and
vary $\alpha$. The observed bias tightly constrains the scatter in the
$L$--$M$ relation (White et al.~2008). In particular, if $F=1$,
$\gamma=1$, the constraints show that $\Delta<0.1$ at the 2-$\sigma$
level, which is tighter than the local relation between black-hole
mass and velocity dispersion. The clustering and density data are not
consistent with $F=1$ at the 1-sigma level for any of the parameter
sets shown.  Since the scatter $\Delta$ must be finite, the contours
therefore suggest that $F>1$, even at the highest possible duty-cycle.

\begin{figure*}
\centerline{\epsfxsize=4.9in \epsfbox{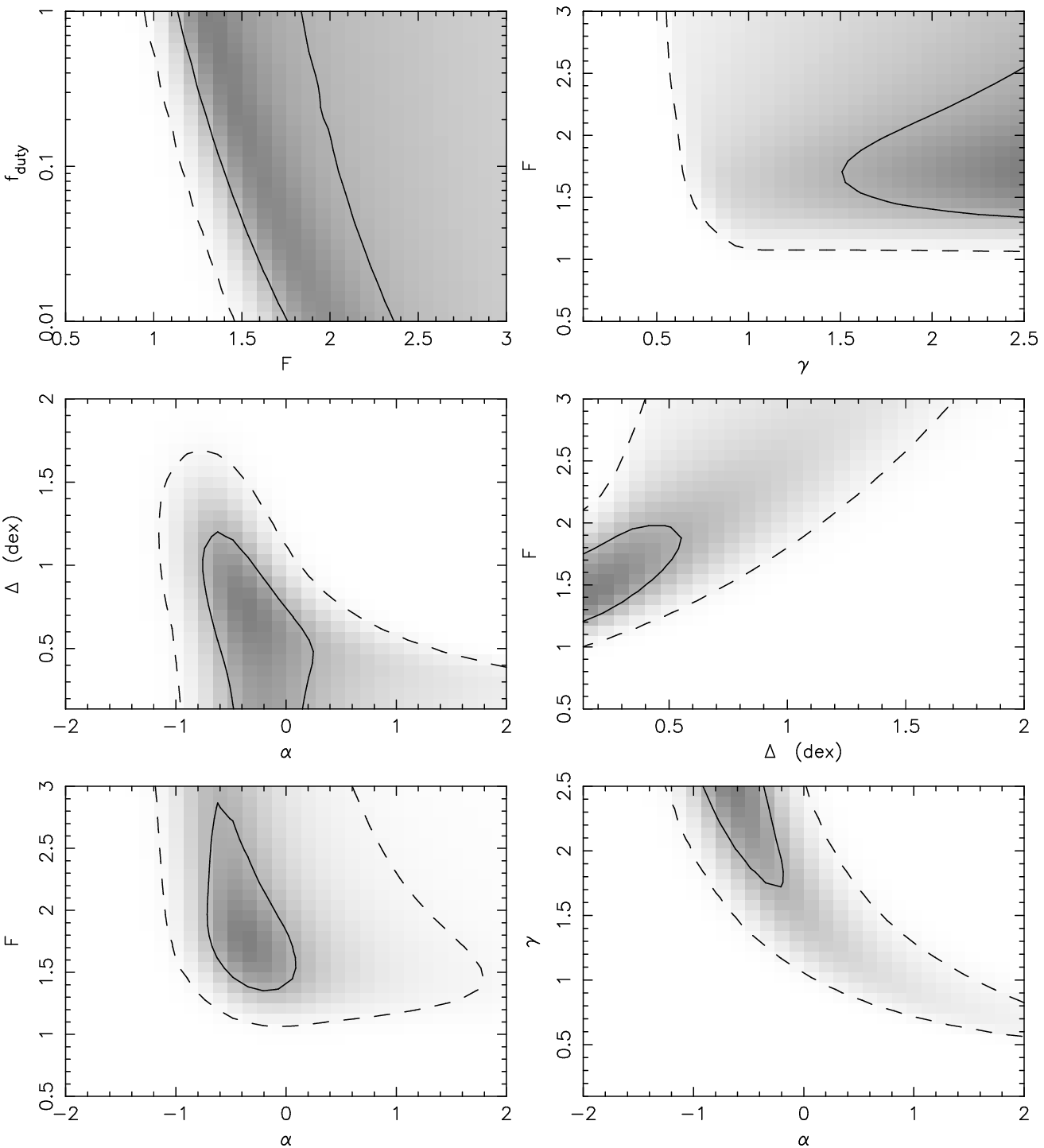}} 
\caption{Contours of the joint likelihood distributions for $f_{\rm
duty}$ and $F$ (top left panel), for $\gamma$ and $F$ (top right
panel), for $\Delta$ and $\alpha$ (central left panel), for $F$ and
$\Delta$ (central right panel), for $F$ and $\alpha$ (lower left
panel) and for $\gamma$ and $\alpha$ (lower right panel). Contours are
shown at 60\% (solid) and 7\% (dotted) of the maximum likelihood. See
text for details on the assumed prior probability distributions.}
\label{fig3}
\end{figure*}

In addition to the observed bias, our model is constrained by the
observed luminosity function slope. In Figures~\ref{fig1} and
\ref{fig2} we also show the region of parameter space that is
consistent with the constraint from the observed luminosity function
slope (grey scale). The predicted slope is sensitive to the value of
$\gamma$. As a result, the values of $f_{\rm duty}$ and $\Delta$ that
are consistent with the observed slope are also sensitive to
$\gamma$. Viable models must have overlapping regions of parameter
space that satisfy the different
constraints. Figures~\ref{fig1}-\ref{fig2} show that unless $\gamma$
or $\alpha$ are large, there are no combinations of $f_{\rm duty}$ and
$\Delta$ that are even marginally consistent with the constraints of
observed bias and luminosity function slope for $F=1$.

\subsection{Joint Likelihood distributions for pairs of parameters}

To quantify the above conclusions we next calculate the joint
likelihoods for pairs of parameters, and the a'posteriori probability
distributions for individual parameters.

The likelihood for the parameter set $(\Delta,\gamma,f_{\rm
duty},F,\alpha)$ is
\begin{eqnarray}
\nonumber
\mathcal{L}_{\Delta,\gamma,f_{\rm duty},F,\alpha}&=&\\
&&\hspace{-20mm}\exp{\left(-\frac{1}{2}\left[\left(\frac{F\langle b\rangle-14.2}{1.4}\right)^2 + \left(\frac{\beta-2.58}{0.23}\right)^2\right]\right)}.
\end{eqnarray}
We then obtain marginalised likelihoods by integrating over the
remaining parameters. For example, the joint likelihood for $\Delta$
and $f_{\rm duty}$ is
\begin{eqnarray}
\nonumber
\label{likelihood}
\mathcal{L}_{\Delta,f_{\rm duty}}&\propto&\int_{\alpha_{\rm
min}}^{\alpha_{\rm max}}d\alpha\int_{\gamma_{\rm min}}^{\gamma_{\rm
max}}d\gamma\int_{F_{\rm min}}^{F_{\rm max}}dF\,
\mathcal{L}_{\Delta,\gamma,f_{\rm duty},F,\alpha}\\
&&\hspace{20mm}\times\frac{dP_{\rm prior}}{d\alpha}\frac{dP_{\rm
prior}}{d\gamma}\frac{dP_{\rm prior}}{dF},
\end{eqnarray}
where $\alpha_{\rm min}$ and $\alpha_{\rm max}$ are the bounding
values on the finite region of the prior probability distribution
$dP_{\rm prior}/d\alpha$ for the variable $\alpha$, and so on. We
assume constant prior probability distributions for $\alpha$, $\Delta$
and $\gamma$. For $F$ and $t_{\rm duty}$, which are fractional
quantities, we assume  prior probability distributions that are flat in
the logarithm of these quantities. We assume the prior probability to
be non-zero within the following ranges: $\alpha_{\rm min}=-2$,
$\alpha_{\rm max}=2$; $\gamma_{\rm min}=0$, $\gamma_{\rm max}=2.5$;
$\Delta_{\rm min}=0$, $\Delta_{\rm max}=2$; $F_{\rm min}=0.5$, $F_{\rm
max}=3$; $f_{\rm duty,min}=0.01$, $f_{\rm duty,max}=1$. While some of
our quantitative results are sensitive to the values of these limits,
our primary qualitative conclusions are robust to the choice of the
prior probability distributions and their ranges.

Figure~\ref{fig3} shows contours of the joint likelihood distributions
for $f_{\rm duty}$ and $F$ (top left panel), for $\gamma$ and $F$
(top right panel), for $\Delta$ and $\alpha$ (central left panel), for
$F$ and $\Delta$ (central right panel), for $F$ and $\alpha$ (lower
left panel) and for $\gamma$ and $\alpha$ (lower right
panel). Contours are shown at 60\% and 7\% of the maximum
likelihood. Since the number of free parameters exceeds the number of
observables, a unique model cannot be found. Nevertheless, the results
of Figure~\ref{fig3} illustrate that despite some degeneracies,
several interesting statements can be made regarding the allowed
parameter values.

Our main results are as follows. First, the parameter $\alpha$ is
constrained to lie within a finite range centered around
$\alpha\approx -0.3$. In addition, lower limits can be placed on the
value $\gamma$, which we find to be larger than unity. The value of
the scatter is restricted to be $\Delta\la1$ dex. This constraint is
much less stringent than the conclusion of White, Martini \&
Cohn~(2008). The difference can be traced to the fact that White et
al. (2008) assumed the halo bias to be equal to the observed bias. We
find that degeneracies prevent the data from imposing an upper limit
on $\gamma$ or a lower limit\footnote{Note that the value of $\Delta$
  would be expected lie in excess of the scatter of $\sim0.3$dex in
  the local relation between velocity dispersion and black-hole
  mass. Inclusion of this constraint as a prior on $\Delta$ tightens
  the constraints on other parameters but does not change our
  qualitative conclusions.}  on $\Delta$ (although the latter cannot
be negative). We find that the duty cycle is unconstrained by these
observations. This result differs from previous studies (e.g. Martini
\& Weinberg~2001) because the parameter $F$ removes the direct
connection between halo bias and halo mass (and hence halo
density). The value of $F$ is constrained to be larger than unity.
This is our most important conclusion, and we will return to its
implications in \S~\ref{discussion}.

\subsection{Constraints on individual parameters}

\begin{figure*}
\centerline{\epsfxsize=6.5in \epsfbox{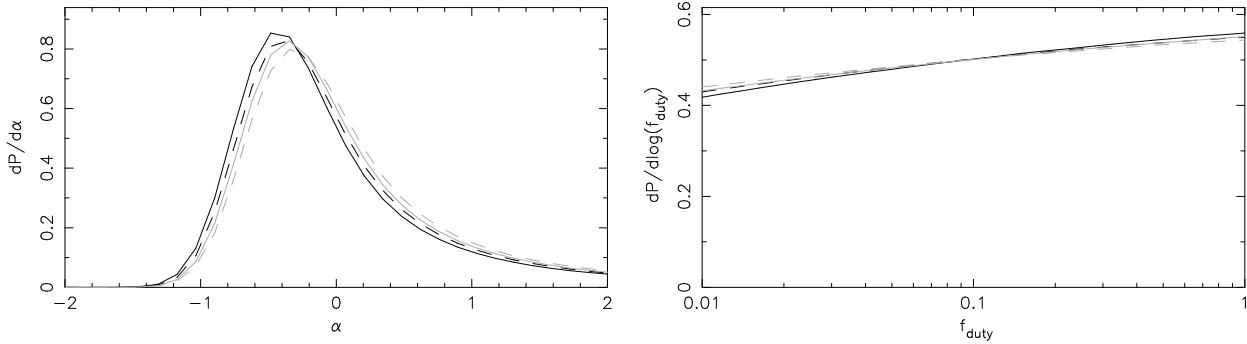}} 
\caption{Marginalised, a'posteriori differential probability
distributions for $\alpha$ (left) and $f_{\rm duty}$ (right). The
black and grey lines correspond to calculations performed using the
Press-Schechter and Sheth-Tormen formalisms respectively. The solid
(dashed) lines do (do not) include the correction factor of 2 in the
quasar density (see text for details).}
\label{fig4}
\end{figure*}

\begin{figure*}
\centerline{\epsfxsize=6.5in \epsfbox{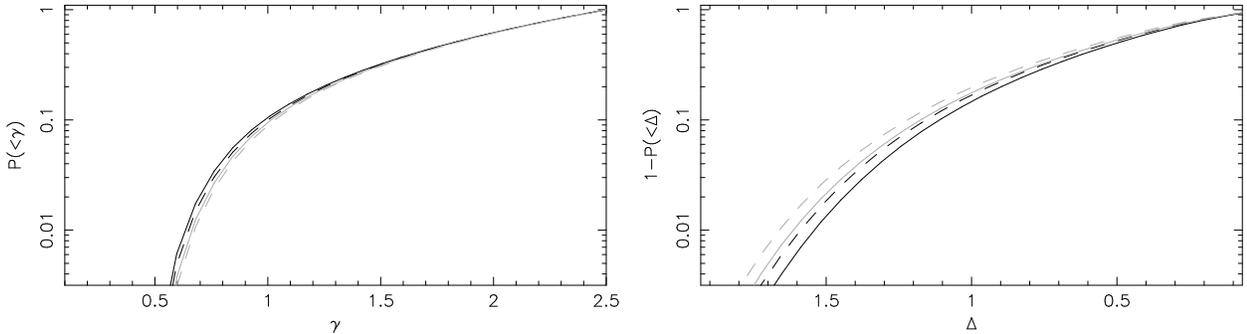}} 
\caption{Marginalised, a'posteriori cumulative probability
distributions for $\gamma$ (left) and $\Delta$ (right). The black and
grey lines correspond to calculations performed using the
Press-Schechter and Sheth-Tormen formalisms respectively. The solid
(dashed) lines do (do not) include the correction factor of 2 in the
quasar density.}
\label{fig5}
\end{figure*}
\begin{figure*}
\centerline{\epsfxsize=6.5in \epsfbox{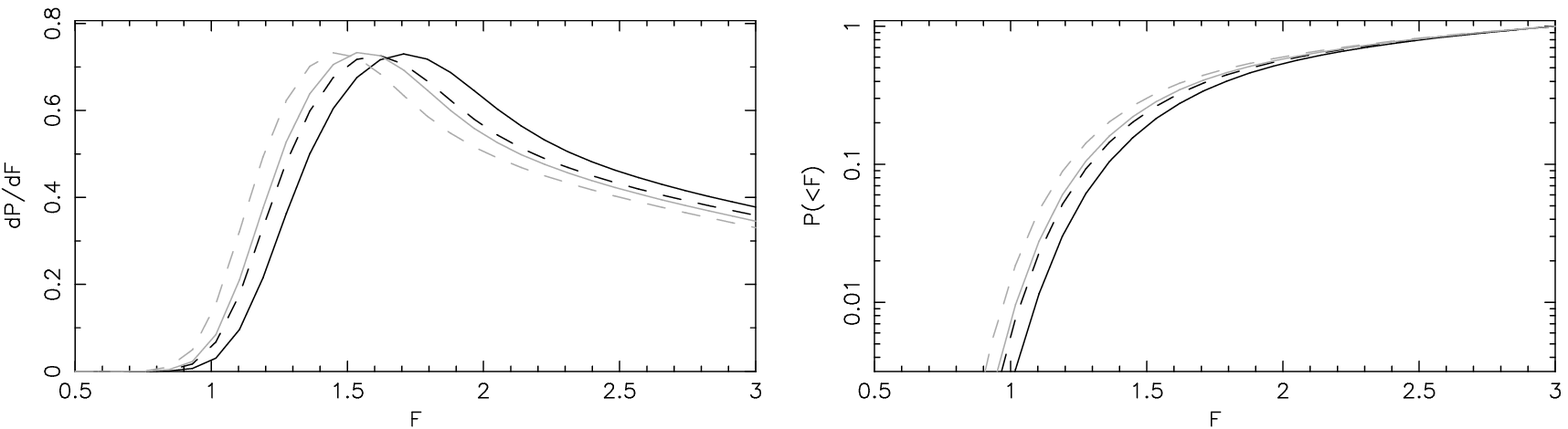}} 
\caption{Marginalised, a'posteriori differential (left) and cumulative
(right) probability distributions for $F$. The black and grey lines
correspond to calculations performed using the Press-Schechter and
Sheth-Tormen formalisms respectively. The solid (dashed) lines do (do
not) include the correction factor of 2 in the quasar density.}
\label{fig6}
\end{figure*}
\begin{figure*}
\centerline{\epsfxsize=6.5in \epsfbox{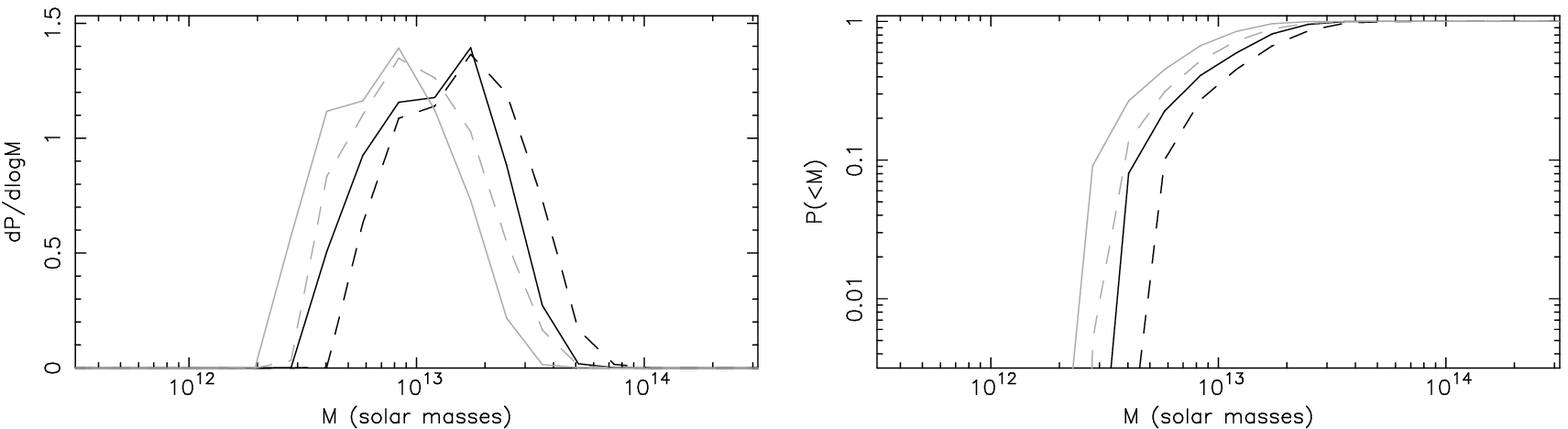}}
\caption{ Marginalised, a'posteriori differential (left) and
cumulative (right) probability distributions for halo mass $M$. The
black and grey lines correspond to calculations performed using the
Press-Schechter and Sheth-Tormen formalisms respectively. The solid
(dashed) lines do (do not) include the correction factor of 2 in the
quasar density.}
\label{fig7}
\end{figure*}
By analogy with equation~(\ref{likelihood}), the joint likelihood
function $\mathcal{L}_{\Delta,\gamma,f_{\rm duty},F,\alpha}$ can be
marginalised over all remaining parameters to find the likelihood for
a single variable. For example, the likelihood for $\Delta$ is
\begin{eqnarray}
\nonumber \mathcal{L}_{\Delta}&\propto&\int_{f_{\rm duty,min}}^{f_{\rm
duty,max}}df_{\rm duty}\int_{\alpha_{\rm min}}^{\alpha_{\rm
max}}d\alpha\int_{\gamma_{\rm min}}^{\gamma_{\rm
max}}d\gamma\int_{F_{\rm min}}^{F_{\rm max}}dF\\
&&\hspace{1mm}\mathcal{L}_{\Delta,\gamma,f_{\rm
duty},F,\alpha}\frac{dP_{\rm prior}}{d\alpha}\frac{dP_{\rm
prior}}{df_{\rm duty}}\frac{dP_{\rm prior}}{d\gamma}\frac{dP_{\rm
prior}}{dF}.
\end{eqnarray}
The a'posteriori probability density for $\Delta$ becomes
\begin{eqnarray}
\nonumber
\frac{dP}{d\Delta}\propto\mathcal{L}_{\Delta}\frac{dP_{\rm prior}}{d\Delta},
\end{eqnarray}
with the normalisation,
\begin{eqnarray}
\nonumber
\int_{\Delta_{\rm min}}^{\Delta_{\rm max}}\frac{dP}{d\Delta}d\Delta=1.
\end{eqnarray}
The corresponding a'posteriori cumulative probability is
\begin{eqnarray}
\nonumber
P(<\Delta)=\int_{\Delta_{\rm min}}^\Delta d\Delta^\prime\frac{dP}{d\Delta^\prime}.
\end{eqnarray}

Figure~\ref{fig4} shows the differential distributions $dP/d\alpha$
and $dP/df_{\rm duty}$. The data constrains a value of $\alpha\approx
-0.4\pm0.5$, while the value of $f_{\rm duty}$ remains
unconstrained. The left-hand panel of Figure~\ref{fig5} shows the
cumulative distribution for $\gamma$. This distribution shows that
$\gamma$ must be greater than unity at a statistical confidence of
90\%. Similarly, the right-hand panel shows the value of 1 minus the
cumulative probability distribution for $\Delta$. At 90\% confidence,
the scatter can only be shown to be below $\Delta=1.1$ (in difference
from White et al.~2008). Finally, Figure~\ref{fig6} plots the
differential (left, $dP/dF$) and cumulative [right, $P(<F)$]
distributions for $F$. The most likely value is at $F\approx 1.7$. The
probability drops rapidly towards low values of $F$ and we find that
$F>1.1$ with high statistical confidence (99\%).

As noted above, our quantitative results are sensitive to the prior
probability distributions because the number of free parameters is
larger than the number of constraints. In particular, we find that the
data does not constrain the upper limit of the parameters $\gamma$ or
$F$. As a result, our quantitative results are sensitive to the values
chosen for the parameters of $\gamma_{\rm max}$ and $F_{\rm
max}$. However, if we increase the values of $\gamma_{\rm max}$ and
$F_{\rm max}$, permitting a greater volume of parameter space
a'priori, then our constraints on the lower limits for $\gamma$ and
$F$ are improved. Therefore, our qualitative conclusions are not
sensitive to the assumed prior probability distribution. In summary,
we find constraints on individual parameters of $\alpha\approx
-0.4\pm0.5$, $\gamma>1$ (at a statistical confidence of 90\%),
$\Delta<1.1$ (90\%) and $F>1.1$ (99\%).

\subsection{Constraints on host halo mass}

Finally, we are able to use the above constraints to assess the range
of halo masses that are consistent with the clustering and luminosity
function data. In our formalism the halo mass is not a free
parameter. Rather, for each parameter set $\Delta,\gamma,f_{\rm
duty},F,\alpha$, there is a halo mass $M_{\Delta,\gamma,f_{\rm
duty},F,\alpha}$, which corresponds to the observed quasar
density. Since our formalism includes scatter in the $L$--$M$
relationship, this mass is defined as the mean mass $M$ at luminosity
$L$ (equation~\ref{M-L}).  This parameter combination has likelihood
$\mathcal{L}_{\Delta,\gamma,f_{\rm duty},F,\alpha}$. The probability
distribution for $M$ can therefore be computed from
\begin{eqnarray}
\nonumber \frac{dP}{dM}&\propto&\\
\nonumber&&\hspace{-15mm}\int_{\Delta_{\rm min}}^{\Delta_{\rm
max}}d\Delta\int_{f_{\rm duty,min}}^{f_{\rm
duty,max}}df_{\rm duty}\int_{\alpha_{\rm min}}^{\alpha_{\rm
max}}d\alpha\int_{\gamma_{\rm min}}^{\gamma_{\rm
max}}d\gamma\int_{F_{\rm min}}^{F_{\rm max}}dF\\
\nonumber&&\hspace{-10mm}\mathcal{L}_{\Delta,\gamma,f_{\rm
duty},F,\alpha}\frac{dP_{\rm prior}}{d\Delta}\frac{dP_{\rm
prior}}{df_{\rm duty}}\frac{dP_{\rm prior}}{d\alpha}\frac{dP_{\rm prior}}{d\gamma}\frac{dP_{\rm
prior}}{dF}\\
&\times&\delta(M-M_{\Delta,\gamma,f_{\rm duty},F,\alpha}),
\end{eqnarray}
with corresponding cumulative distribution
\begin{equation}
P(<M)=\int_M^\infty dM^\prime \frac{dP}{dM^\prime}.
\end{equation}
We plot these distributions in Figure~\ref{fig7}. The host halo mass
for the quasars in the high redshift sample is
$M\sim10^{13\pm0.5}M_\odot$. In our fiducial model, the halo mass is
restricted to be larger than $\sim3\times10^{12}M_\odot$ (99\%). This
mass range is comparable to the one ($M\ga5\times10^{12}M_\odot$)
quoted by Shen et al.~(2007). However, note that here $M$ refers to
the halo mass corresponding to $L$ at the mean of the $L$--$M$
relation. The possibility of a large intrinsic scatter means that many
host halos will have a significantly smaller mass.

\subsection{Theoretical uncertainties in the bias}
\label{systematic}
Thus far we have presented a fiducial model using the Sheth-Tormen
forms for the clustering bias (Sheth, Mo \& Tormen~1999) and the
mass-function (Sheth \& Tormen~1999). However as demonstrated in
White, Martini \& Cohn~(2008), the bias and density are sufficiently
large as to make any quantitative conclusions sensitive to the
detailed predictions of the Sheth-Tormen model. A discussion
of the various analytic models and their comparison to numerical
simulation is provided by White, Martini \& Cohn~(2008), who suggest
that the Press-Schechter and Sheth-Tormen formalisms should bracket
the expected level of theoretical uncertainty. Following this
assertion, we also present results using the corresponding
Press-Schechter formulation for the mass function
(Press-Schechter~1974) and clustering bias (Mo \& White~1996) to 
assess the level of theoretical systematic uncertainty in our
conclusions.

In addition to the adopted model for the statistics of the halo
population, the conclusions will also be sensitive to the value of the
measured space density of quasars. Shen et al.~(2007) pointed out that
the space density is underestimated by a factor of $\sim 2$ for the
fit of Richards et al.~(2006). Following White, Martini \& Cohn~(2008)
we have used that correction factor in our analysis above. For
comparison, we also repeat our analysis using the uncorrected value of
$N=0.35\times10^{-7}$Mpc$^{-3}$.

We therefore compare a set of four calculations, using combinations of
the Sheth-Tormen or Press-Schechter halo statistics with the quasar
density computed with or without the correction factor. In
Figures~\ref{fig4}-\ref{fig7} we show probability distributions for
the three additional cases together with our fiducial model. The
constraints on $\gamma$ and $\alpha$ are unaffected by these model
changes since these are not directly related to the numbers of
observed quasars. The lower density of quasars and the larger
predicted bias of the Press-Schechter formalism (and their
combination) each yield weaker constraints on $\Delta$, $F$ than our
fiducial model. However the systematic effect is mild, with limits of
$\Delta<1.1$ (85\%), and $F>1.1$ (98\%). We find that the minimum halo
mass is effected by a factor of 2 depending on the assumed
combination.

\section{Discussion}
\label{discussion}

We find that the value of $F$ is constrained to be in excess of unity,
i.e. the observed bias is in excess of the halo bias. This could be
interpreted as evidence that mergers are responsible for the quasar
activity. We also find that $\gamma>1$, implying that the luminosity
to halo mass ratio increases with halo mass. Under the assumption that
the Eddington ratio of bright quasars is nearly constant (as argued by
Kollmeier et al.~2006), models of feedback limited growth through
energy deposition (Haehnelt, Natarajan \& Rees 1998; Wyithe \&
Loeb~2003) predict a value of $\gamma=5/3$.  If we adopt $\gamma=5/3$,
then the preferred range of $F$ (marginalised over other parameters)
is $1.5\la F\la2$. Interestingly, $F\sim1.5$ is approximately the
factor by which Furlanetto \& Kamionkawski~(2006) estimated mergers to
increase the observed bias over the value of halo bias at high values
of $M$. In our analysis we have allowed $F$ to be a free
parameter. This approach has the promise of constraining the merger
process responsible for the triggering of quasar activity. For
example, Furlanetto \& Kamionkawski~(2006) demonstrated that if close
pairs lead to mergers, then mergers are expected to be clustered
differently than individual halos. The magnitude of this difference is
dependent on the clustering model, and so its measurement will
constrain fundamental properties of the merger process.

Several anomalies have been previously noticed in the unexpectedly
high clustering of systems that may be merger driven. For example, the
clustering amplitude of Lyman-break galaxies (Adelberger et al.~2005)
implies a host halo mass of $\sim10^{12}M_\odot$. However at this
mass, the density of Lyman-break galaxies then implies that that the
duty-cycle of vigorous star formation must be unity in all
halos. Moreover, kinematic measurements of these galaxies imply a
significantly smaller mass of $\sim10^{11}M_\odot$ (Pettini et
al.~2001; Erb et al.~2003). The tension between the different
estimates of host mass could be alleviated if the intense star
formation in Lyman-break galaxies is triggered by mergers, and if
mergers in turn increase the observed bias (Furlanetto \&
Kamionkowski~2006). In a recent paper, Wake et al.~(2008) demonstrated
that a radio-loud sub-sample of Luminous Red Galaxies at $z\sim0.55$
had a clustering amplitude that is in excess of the radio-quite
sub-sample, despite the optical properties being identical. Wake et
al.~(2008) interpret this as a larger host halo mass for radio loud
objects. However, it might also be appealing to assign the excess bias
to a merger origin of accretion activity.

Of course there are alternative explanations for each of these
observations. For example, dynamical measurements may only be
accessing the central component of the host halos of Lyman break
galaxies (Cooray 2005). In the case of radio loud Luminous Red
Galaxies, the high bias could originate from the high bias of X-ray
clusters (more massive halos), of which the Luminous Red Galaxies
would be the central member. Additionally, since clustering statistics
have been shown to depend on the halo formation history (Gao, Springel
\& White~2005; Croton, Gao \& White~2007; Wetzel et al.~2007), the age
or structural properties the galaxy host could also effect the
observed clustering. However in each case it is clear that
the possibility of modification of the observed
bias should be taken into account in interpreting
clustering data for many types of galaxies, such as starburst galaxies
or radio galaxies.

\begin{figure}
\centerline{\epsfxsize=2.6in \epsfbox{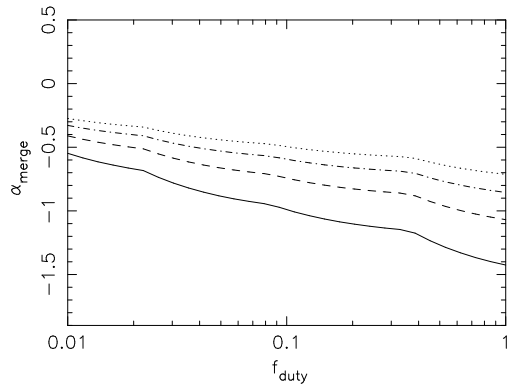}}
\caption{The theoretically expected value of $\alpha_{\rm merge}$ as a
  function of quasar duty cycle (see text for details). Curves are
  plotted for four values of $\gamma=1$, 4/3, 5/3, and 2 (bottom to
  top).  }
\label{fig8}
\end{figure}

The value of $\alpha$ is also well constrained. Since the constraint
of $F>1$ may suggest that quasar activity is triggered by mergers, it is
interesting to calculate the predicted value of $\alpha$ using a model
where the total duty cycle (i.e. summing over all quasar episodes in a
single host) as a function of luminosity is proportional to the rate
of major mergers. We use the extended Press-Schechter formalism to
evaluate the rate of mergers of halos with mass $M_1$ per unit mass
with halos of mass $M_2$, $\left.d^2N/dM_1dt\right|_{M_2}$ (Lacey \&
Cole~1993). The duty-cycle is then proportional to
$M_1\left.d^2N/dM_1dt\right|_{M_2}t_{\rm lt}H$, where $t_{\rm lt}$ is
the lifetime of a single merger driven quasar episode and $H^{-1}$ is
the Hubble time at the redshift of interest (Wyithe \&
Loeb~2003). Assuming that $t_{\rm lt}$ is not a function of mass (as
expected if the quasar episode is related to timescales like the
black-hole doubling time or the dynamical time of its host galaxy), we
find
\begin{equation}
\alpha_{\rm merge} \approx
\frac{1}{\gamma}\frac{d}{d\log{M}}\left[\log{\left(M\left.\frac{d^2N}{dMdt}\right|_{M}\right)}\right],
\end{equation}
where we have assumed major mergers with $M_1\sim M_2=M$, and the term
$1/\gamma$ originates in the conversion from $d\log{M}$ to $d\log{L}$
for consistency with the definition of $\alpha$. The resulting value
of $\alpha_{\rm merge}$ as a function of duty cycle is plotted in
Figure~\ref{fig8} for four values of $\gamma=1$, 4/3, 5/3, and 2
(bottom to top).  While a value of $\gamma=5/3$ applies to feedback
through energy deposition, models where the quasar drives the gas
outflow from its host galaxy through momentum deposition have
$\gamma=4/3$ (e.g. Silk \& Rees 1998; Murray, Quataert \& Thompson 
2005; King~2003).
Here, the value of $\alpha_{\rm merge}$ has been computed at
a halo mass for which the number density times the duty cycle is equal
to the observed quasar density. We find that for this model we would
expect a value of $-1\la\alpha_{\rm merge}\la-0.3$, with larger values
of $\gamma$ indicating less dependence of duty cycle on
luminosity. These values are consistent with our constraints on
$\alpha$.

We note that the scatter is only constrained to satisfy $\Delta\la1$
dex. This range is significantly larger than the tight upper limit of
$\sim0.3$ dex found in the analysis of White et al.~(2008). The
difference can be traced to the greater freedom in our model
associated with: {\it (i)} the relation between luminosity and halo
mass through $\gamma$; {\it (ii)} the relation between quasar density
and halo mass through $\alpha$; and {\it (iii)} the allowance for the
observed clustering bias to not directly reflect the underlying host
mass for merger-driven systems. We do find that models assuming $F=1$,
$\gamma=1$ (as considered by White, Martini \& Cohn~2008) reflect the
small upper limit of $\Delta \sim0.3$ dex. However in addition to this
low value of scatter being disfavoured in our more general model by
the joint constraints from clustering and luminosity function, we
would not expect such a tight relationship on observational
grounds. In particular, an upper limit of 0.3 dex implies a scatter that is smaller than the
scatter in the black hole--velocity dispersion relation measured
locally ($\sim0.3$ dex; Tremaine et al.~2002).  However the relation
between quasar luminosity and halo mass must have several additional
sources of scatter, including in the relation between halo mass and
bulge velocity dispersion, and between black hole mass and quasar
luminosity. Hopkins et al.~(2007) have investigated the scatter in
relations between black-hole mass and galaxy properties including the
bulge velocity dispersion. They argue that the tightness of these
relations implies that any connection between black-hole mass and halo
mass must be incidental. Wyithe \& Loeb~(2005) have modeled the
black-hole -- halo mass relation, and have shown that the scatter in
the black-hole velocity dispersion relation combined with the
statistical properties of dark-matter halos (Bullock et al.~2001)
implies a scatter of at least 0.5 dex at the redshifts of
interest. Moreover, any relation between luminosity and halo mass must
also include a factor to account for the fraction of the Eddington
rate at which the quasar accretes. Kollmeier et al.~(2006) found a
scatter in this distribution of 0.3dex.

Finally, we note that the quasar duty-cycle is not constrained in this
model. This is in difference to previous results (Haiman \& Hui~2001;
Martini \& Weinberg~2001; Shen et al~2007; White et al.~2008), which have
constrained the quasar lifetime from clustering. The difference
originates from the fact that our model does not assume the one to one
correspondence between observed clustering amplitude and halo mass
($F=1$). This generalisation, motivated by the theoretical expectation
that merging systems are likely to be more clustered than their
isolated counterparts at fixed halo mass, means that a large
clustering amplitude can be obtained by more common, lower mass halos,
allowing the observed density to be achieved with a smaller
duty-cycle. Moreover, Figure~\ref{fig3} demonstrates that even if a
value of $F\sim1.5$ (Furlanetto \& Kamionkowski 2006) is adopted, the
duty-cycle still remains unconstrained. This is because the large
value of $F$ allows for reproduction of the data over a wide range of
possible values for the scatter ($\Delta$), which in turn allows the
model to predict a broad range of values for the halo abundance.

In our analysis we have included the slope of the quasar luminosity
function as an additional constraint to the quasar density and
clustering. Before concluding we note that the addition of this
constraint does not effect quantitative conclusions regarding $F$ or
$\Delta$, which are derived from comparison of the halo number density
and bias to the observed density and bias. We have repeated our
analysis without the constraint on the luminosity function slope. We
find that the addition of this constraint allows a limit to be placed on $\alpha$
and tightens the constraint on $\gamma$.

\section{conclusion} 

The large amplitude of the observed clustering among $z\sim4$ quasars,
combined with their density and luminosity function slope, strongly
constrains the relationship between quasar luminosity and halo mass at
high redshifts. We have modelled these observables using the extended
Press-Schechter formalism, combined with a mean quasar luminosity
($L$) -- halo mass ($M$) relation of the form $L\propto M^\gamma$.  We
assume an intrinsic scatter (in dex) around the mean relation of
$\Delta$ at a fixed halo mass. We find that the observed clustering
amplitude and luminosity function slope cannot be simultaneously
reproduced unless {\bf both} of the following conditions hold:

\begin{itemize}

\item The value of $\gamma$ is greater than unity, implying that
quasar luminosity is not a linear function of host mass.

\item The observed clustering amplitude is in excess of that expected
from dark matter halos ($F>1$), possibly implying that the observed
bias is boosted because mergers trigger quasar activity.

\end{itemize}

The latter constraint on $F$ is particularly strong. We find $F>1.1$ at
the 99\% level.  On the other hand, we find that the scatter in the
$L-M$ relation can be constrained only to a value smaller than
$\sim1$ dex (in difference from the recent study of White et al.~2008). We
find that because of the weak constraint on the scatter the mean host mass
can be only weakly constrained, to within an order of magnitude around
$\sim10^{13}M_\odot$. Importantly, and for the same reason, we find
that clustering data combined with quasar densities does not constrain
the quasar lifetime, as had been suggested previously (Martini \&
Weinberg~2001; Haiman \& Hui~2001).

Interestingly, the literature already contains a physically motivated
model which is consistent with all of the above constraints. A
scenario in which galaxy mergers drive accretion onto a central black
hole, which then radiates near its Eddington luminosity until it grows
sufficiently to deposit (with a $\sim 5\%$ efficiency) the binding
energy of the surrounding gas over a dynamical time, yields an
$L$--$M$ relation with a value of $\gamma=5/3$ (Wyithe \&
Loeb~2003). For this value of $\gamma$, our constraints favor a value
of $1.5\la F\la2$, which is consistent with estimates of the
expected enhancement in clustering bias for merger driven activity of
$F\sim1.5$ (Furlanetto \& Kamionkawski~2006).

In summary, the clustering statistics combined with the luminosity
function of high redshift quasars, provide compelling evidence for a
scenario of merger-driven quasar activity, with black-hole growth that
is limited by feedback. The resulting luminosity--halo mass relation
implies that the luminosity to halo mass ratio increases with halo
mass as expected in theoretical models that include feedback limited black
hole growth.

\bigskip

\noindent
{\bf Acknowledgments.} The research was supported by the Australian
Research Council (JSBW), by NASA grants NNX08AL43G and by Harvard
University funds (AL).

\label{lastpage}
\end{document}